\documentclass[12pt]{article}
\usepackage{latexsym,amssymb,amsmath}
\usepackage{graphicx}                   
\begin{document}

\thispagestyle{empty} \thispagestyle{empty}

\title{\bf Reconstruction of the three mechanical material constants of a lossy
fluid-like cylinder from low-frequency scattered acoustic
fields\rm}
\author{Thierry Scotti\thanks {Laboratoire de M\'ecanique et
d'Acoustique, UPR 7051 du CNRS,
 31 chemin Joseph Aiguier, 13009 Marseille, France, wirgin@lma.cnrs-mrs.fr} \and Armand Wirgin
 \thanks{Laboratoire de M\'ecanique et d'Acoustique, UPR 7051 du CNRS,
 31 chemin Joseph Aiguier, 13009 Marseille, France, ogam@lma.cnrs-mrs.fr}.}

\maketitle

%

\begin{abstract}
The inverse medium problem for a circular cylindrical domain is
studied using low-frequency acoustic waves as the probe radiation.
It is shown that to second order in $k_{0}a$ ($k_{0}$ the
wavenumber in the host medium, $a$ the radius of the cylinder),
only the first three terms (i.e., of orders 0, -1 and +1) in the
partial wave representation of the scattered field are
non-vanishing, and the material parameters enter into these terms
in explicit manner. Moreover, the zeroth-order term contains only
two of the unknown material constants (i.e., the real and
imaginary parts of complex compressibility of the cylinder
$\kappa_{1}$) whereas the $\pm 1$ order terms contain the other
material constant (i.e., the density of the cylinder $\rho_{1}$).
A method, relying on the knowledge of the totality of the far-zone
scattered field and resulting in explicit expressions for
$\rho_{1}$ and $\kappa_{1}$, is devised and shown to give
highly-accurate estimates of these quantities even for frequencies
such that $k_{0}a$ is as large as 0.1.
\end{abstract}
\newpage
\newpage
\section{Introduction}\label{s1}
Retrieving  the mechanical material constants (e.g., elastic
moduli, density) of a material, either by direct measurements of
these quantities, or by measurements of other variables from which
the material constants can be derived with the help of suitable
models, is one of the central problems in materials science. When
the specimen is a solid, the elastic moduli are determined either
by the usual static methods or by dynamic methods involving the
inversion of data relative to  resonant frequencies and/or mode
shapes of vibrations excited, for instance by percussive forces
\cite{CaAd87}, \cite{Va90}, \cite {GrLe92}, \cite{NeWo92}, or
relative to velocities and attenuations for ultrasonic wave probe
radiation \cite{KrKr69}. Ultrasound methods can also be employed
for fluids, or fluid-like materials \cite{KrKr69}, \cite{RoGr86}.

Another class of material characterization methods, called {\it
resonance spectroscopy} \cite{AlDe86} combines the underlying
principles of vibratory resonances with an acoustic excitation of
the specimen. This technique has also been employed to determine
the refractive index of beads by means of laser irradiation
\cite{ChRa83}. This class of techniques differs from the previous
ones in that it relies on measurements of the {\it wavefield
diffracted by the specimen}, and appeals to a quite intricate
theory relating the resonances to coefficients computed from the
diffracted field for estimating the material parameters (it can
also be employed for estimating the geometrical parameters of the
specimen \cite{ChRa83}, \cite{AlDe86}, \cite{DaSo81}). This theory
is only feasible for specimens having simple geometry (e.g.,
spherical, circular cylindrical, plate-like). A simple specimen
geometry is also required in the standard vibration-resonance and
velocity-attenuation methods if absolute quantitative
characterizations are aimed at.

During the last 25 years, another materials-characterization
method has been developed which can be termed {\it wavefield
imaging}. The underlying idea is: i) acquire measurements of the
field scattered from a specimen at a series of locations in space
arising either from several monochromatic probe fields, and/or
from a pulse-like probe field, and ii) retrieve from these
measurements an {\it image} of the specimen (i.e., a spatial map
of some material characteristic, such as wavespeed or attenuation)
with the help of a suitable model of the specimen/wave
interaction. Insofar as there is a sharp difference between the
material properties of the specimen and those of the host medium,
this method also gives a picture of the geometry (location,
orientation, size and shape) of the specimen. When, as is often
the case, the information relating to the material constants of
the specimen is not reliable, only the geometrical information can
been exploited (this is called {\it qualitative wavefield
imaging}; otherwise it is called {\it quantitative wavefield
imaging}). For instance, computerized diffraction tomography,
making use of a model appealing either to the Rytov or Born
approximations of the specimen/wave interaction, is a qualitative
wavefield imaging technique except for specimens whose properties
differ only slightly from those of the host medium (this is
fortunately the case in biological imaging applications)
\cite{RoGr86}, \cite{To94}, \cite{DeLe97}, \cite{De99}. It has
been suggested \cite{To94}, \cite{DeLe97}, \cite{De99} that one of
the reasons why Born-based techniques do not furnish reliable
estimates of the material properties (notably the wavespeed, in
specimens assumed to be lossless and surrounded by a host medium
which is also lossless and has the same density as that of the
specimen), is that data relating to low-frequency probe radiation
was either not available or not used in the inversion algorithm.

The importance of disposing of multi-frequency (and, in
particular, low frequency) data is increasingly recognized as the
key to success for material characterization in wavefield imaging
techniques such as the distorted Born method \cite{HaEb97},
\cite{TiBe01}, the modified Born and modified gradient methods
\cite{BeTi01}, and the contrast source method \cite{BlAb01}. The
possibility of obtaining a quantitatively-accurate image with
these iterative methods is often dependent on being able to
initialize the algorithm with a plausible image of the object at
the lowest frequency of the probe radiation. More often than not,
this initial image is obtained via the Born approximation, and
since the latter is not accurate for large contrasts (between the
host and the object) of the material constants, the algorithm has
trouble with restoring the right values of the material constants
during the iterative process. Thus, it would be useful to find a
means for obtaining a better estimate of the material constants at
low frequencies in the case of arbitrarily-large contrasts. This
is done herein.

In particular, we shall be concerned with the retrieval of the
three material constants $\Re\kappa_{1}$, $\Im\kappa_{1}$ and
$\rho_{1}$ (wherein $\kappa_{1}=\Re\kappa_{1}+i\Im\kappa_{1}$) of
a generally-lossy fluid object in a lossless fluid host probed by
plane-wave acoustic radiation. The case $\Im\kappa_{1}=0$ of a
lossless material can also be treated. No assumption is made
concerning the contrasts of density and compressibility between
the host and the object. The latter is assumed to be a  circular
cylinder, of known radius $a$. The material constants of the host
medium (in which the probe radiation propagates) are also assumed
to be known, as is known the frequency and incident angle of the
plane wave probe radiation, as well as the scattered acoustic
wavefield in the far zone of the cylinder. The analysis for
recovering the three material parameters of the cylinder is
focused on the case in which the wavelength
($\lambda_{0}=2\pi/k_{0}$, with $k_{0}=\omega/c_{0}$, $c_{0}$ the
speed of bulk waves in the host, and $\omega$ the angular
frequency) of the probe radiation is much larger than the cylinder
radius.
\section{Physical configuration and governing equations}\label{s2}
The scattering body is an infinite cylinder whose generators are
parallel to the $z$ axis in the cylindrical polar coordinate
system $(r,\theta,z)$. The intersection of the cylinder, within
which is located the origin $O$, with the $xOy$ plane defines (see
figure 1):

i) the boundary curve $\Gamma=\{r=\rho(\theta); 0\le \theta < 2\pi
\}$, with $\rho$ a continuous, single-valued function of $\theta$
(further on, we shall take $\Gamma$ to be a circle, i.e.,
$\rho(\theta)=a$),

ii) the bounded (inner) region (i.e., that occupied by the body in
its cross-section plane) $\Omega_{1}=\{r<\rho(\theta); 0\le \theta
< 2\pi \}$,

iii) the unbounded (outer) region $\Omega_{0}=\{r>\rho(\theta);
0\le \theta < 2\pi \}$.
\begin{figure}
[ht]
\begin{center}
  \includegraphics[scale=0.5] {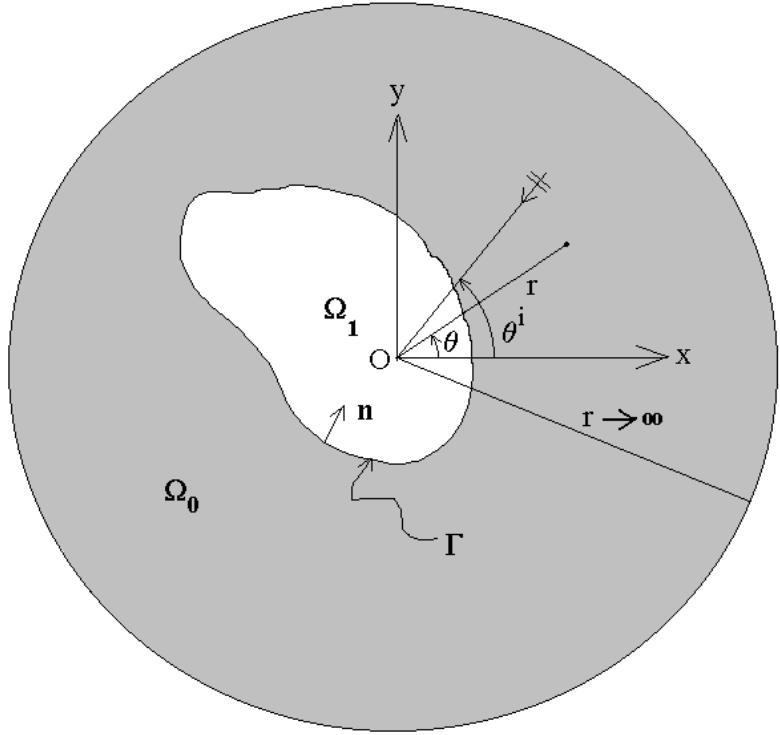}
  \caption{Problem configuration in the $xOy$ plane}
  \label{fullobstacle}
  \end{center}
\end{figure}

It is assumed that $\Omega_{0}$ and $\Omega_{1}$ are filled with
linear, homogeneous, isotropic, time-invariant fluid-like media
$M_{0}$ and $M_{1}$ respectively and that $M_{1}$ is possibly
lossy.

The cylinder is probed by a monochromatic acoustic plane whose
propagation vector lies in the $xOy$ plane. Due to the invariance
of the cylinder and incident field with respect to $z$, the
incident and scattered field is also invariant with respect to
$z$. Let $U$ designate pressure, which, due to the
previously-mentioned invariance, is of the form:
\begin{equation}\label{2.1}
U=U(\mathbf{x},t)~,
\end{equation}
with $\mathbf{x}=(x,y)=(r,\theta)$. This invariance applies also
when superscripts $i$ and $d$ ($i$ for 'incident' and $d$ for
'diffracted') are attached to $U$. It is convenient to associate
$U(\mathbf{x},t)$ with the total field, it being understood that
the latter takes the form $U_{j}(\mathbf{x},t)$ in $\Omega_{j}$
and:
\begin{equation}\label{2.2}
U_{j}(\mathbf{x},t)=U^{i}(\mathbf{x},t)\delta_{j0}+
U_{j}^{d}(\mathbf{x},t)~;~\mathbf{x}\in\Omega_{j}~,
\end{equation}
with $\delta_{jk}$ the Kronecker delta.

We express $U$ by the Fourier transform
\begin{equation}\label{2.3}
U_{j}(\mathbf{x},t)=\int_{-\infty}^{\infty}u_{j}(\mathbf{x},\omega)\exp(-i\omega
t)d\omega~,
\end{equation}
with similar expressions for  $U^{i}$ and $U^{d}$. The
monochromatic, plane-wave nature of the incident field is such
that
\begin{equation}\label{2.4}
u^{i}(\mathbf{x},\omega)=\exp\left [ -ik_{0}r\cos\left (
\theta-\theta^{i}\right ) \right ] ~,
\end{equation}
wherein $\theta^{i}$ designates the incident angle.

The essential task in both the forward and inverse scattering
contexts is to determine
\begin{equation}\label{2.6}
u_{j}(\mathbf{x},\omega)=u^{i}(\mathbf{x},\omega)\delta_{j0}+
u_{j}^{d}(\mathbf{x},\omega)~;~\mathbf{x}\in\Omega_{j}~.
\end{equation}
via the governing equations:
\begin{equation}\label{2.7}
(\Delta
+k_{j}^{2})u_{j}(\mathbf{x},\omega)=0~;~\mathbf{x}\in\Omega_{j}~~,~~j=0,1~,
\end{equation}
\begin{equation}\label{2.8}
\lim_{ r\rightarrow \infty}r^{1/2}(\partial_{r}
-ik_{0})u_{0}^{d}(\mathbf{x},\omega)=0~;~\forall\theta\in[0,2\pi[
~,
\end{equation}
\begin{equation}\label{2.11}
|u_{1}^{d}(\mathbf{x},\omega)|<\infty~;~\mathbf{x}\in\Omega_{1}~,
\end{equation}
\begin{equation}\label{2.12}
u_{0}(\mathbf{x},\omega)-u_{1}(r,\theta,\omega)=0~;~\mathbf{x}\in\Gamma~,
\end{equation}
\begin{equation}\label{2.13}
\alpha_{0}\mathbf{n}\cdot\nabla
u_{0}(\mathbf{x},\omega)-\alpha_{1}\mathbf{n}\cdot\nabla
u_{1}(\mathbf{x},\omega)=0 ~;~\mathbf{x}\in\Gamma~,
\end{equation}
wherein $\mathbf{n}$ is the unit vector normal to $\Gamma$ and:
\begin{equation}\label{2.14}
k_{j}^{2}=\omega^{2}(c_{j})^{-2}~,~~
c_{j}^{2}=(\rho_{j}\kappa_{j})^{-1}~,~~
\alpha_{j}=(\rho_{j})^{-1}~.
\end{equation}
\section{Forward and inverse scattering problems}\label{s4}
The {\it forward scattering problem} (notably for simulating
measured data) is formulated as follows:
\newline
given:

i) the location, shape, size and composition (material properties
) of the scattering body,

ii) the material properties of the host medium $M_{0}$

iii) the incident wavefield (i.e., (\ref{2.4}), as well as the
frequency thereof,
\newline
determine:

the field (i.e., $u_{j}^{d}~;~j=0,1$) scattered by the body at
arbitrary points of space.

The general {\it inverse scattering problem} is formulated as
follows:
\newline
given:

i) the incident wavefield (i.e., (\ref{2.4})), as well as the
frequency thereof,

ii) the material properties of the host medium $M_{0}$

ii) the wavefield in some subregion of $\Omega_{0}$,
\newline
reconstruct:

the location, shape, size and composition of the scattering body.
\\\\
Hereafter, we shall be concerned mostly with the {\it inverse
problem}, and, in particular, with one in which the location, size
and shape of the body are known beforehand, the task being to
reconstruct the {\it composition} of the body.

In fact, the body will be chosen to be a {\it homogeneous cylinder
with center at the origin} $O$ and radius $a$, and we will try to
reconstruct its material properties $\rho_{1}$ and $\kappa_{1}$
from the scattered acoustic field for {\it low-frequency probe
radiation}.
\section{Partial wave expressions of the fields}\label{s5}
A well-known identity \cite{AbSt68}
informs us that the plane-wave probe radiation admits the partial
wave expansion
\begin{equation}\label{5.3}
u^{i}(\mathbf{x},\omega)=\sum_{m=-\infty}^{\infty}
\gamma_{m}J_{m}(k_{0}r)\exp\ ( im\theta\ )~~;~~\forall
\mathbf{x}\in \mathbb{R}^{2}~.
\end{equation}
wherein $J_{m}(~~)$ is the $m$-th order Bessel function and
\begin{equation}\label{5.5}
\gamma_{m}=\exp\ ( -im(\theta^{i}+\pi/2)\ )~.
\end{equation}
By applying the separation of variables technique to (\ref{2.7})
and (\ref{2.8}), it is found that
\begin{equation}\label{6.1}
u_{0}^{d}(\mathbf{x},\omega)=\sum_{m=-\infty}^{\infty}
C_{m}H_{m}^{(1)}(k_{0}r)\exp\ ( im\theta\ )~~;~~\forall
\mathbf{x}\in \Omega_{0}~,
\end{equation}
wherein $H_{m}^{(1)}(~~)$ is the $m$-th order Hankel function. The
application of the same technique to (\ref{2.7}) and (\ref{2.11})
gives
\begin{equation}\label{6.2}
u_{1}^{d}(\mathbf{x},\omega)=\sum_{m=-\infty}^{\infty}
D_{m}J_{m}(k_{1}r)\exp\ ( im\theta\ )~~;~~\forall \mathbf{x}\in
\Omega_{1}~.
\end{equation}

From the transmission conditions (\ref{2.12}) and (\ref{2.13}) we
deduce, due to the orthogonality of the functions
$\{\exp(im\theta)\}$, the fact that $\rho(\theta)=a$ and
$\mathbf{n}\cdot\nabla=-\partial_{r}$ in the present case,
\begin{equation}\label{7.4}
C_{m}=\gamma_{m}\frac{J_{m}(k_{0}a)\dot{J}_{m}(k_{1}a)-
\beta\dot{J}_{m}(k_{0}a)J_{m}(k_{1}a)}
{\beta\dot{H}_{m}^{(1)}(k_{0}a)J_{m}(k_{1}a)-
H_{m}^{(1)}(k_{0}a)\dot{J}_{m}(k_{1}a)}~,
\end{equation}
\begin{equation}\label{7.5}
D_{m}=\gamma_{m}\beta\frac{J_{m}(k_{0}a)\dot{H}_{m}^{(1)}(k_{0}a)-
\dot{J}_{m}(k_{0}a)H_{m}^{(1)}(k_{0}a)}
{\beta\dot{H}_{m}^{(1)}(k_{0}a)J_{m}(k_{1}a)-
H_{m}^{(1)}(k_{0}a)\dot{J}_{m}(k_{1}a)}~,
\end{equation}
wherein: $Z_{m}(\xi)=J_{m}(\xi)$, $Z_{m}(\xi)=Y_{m}(\xi)$ or any
linear combination thereof, knowing that
$H_{m}^{(1)}(\xi)=J_{m}(\xi)+iY_{m}(\xi)$, with $Y_{m}(\xi)$ the
$m$-th order Neumann function), $\dot{Z}(\xi):=dZ(\xi)/d\xi$ and
$\beta=k_{0}\alpha_{0}/k_{1}\alpha_{1}$.
\section{Low-frequency approximation of the scattered field outside
of the body and inversion formulas}\label{s8} We use the formulas
\cite{AbSt68}
\begin{equation}\label{8.1}
\dot{Z}_{m}(\xi)=Z_{m-1}(\xi)-\frac{m}{\xi}\dot{Z}_{m}(\xi)~,~~
Z_{-m}(\xi)=(-1)^{m}Z_{m}(\xi)~,
\end{equation}
 to find:
\begin{equation}\label{8.3}
\dot{Z}_{0}(\xi)=-Z_{1}(\xi)~,
\end{equation}
and
\begin{equation}\label{8.4}
C_{0}=\gamma_{0}\frac{-J_{0}(k_{0}a)J_{1}(k_{1}a)+ \beta
J_{1}(k_{0}a)J_{0}(k_{1}a)} {-\beta
H_{1}^{(1)}(k_{0}a)J_{0}(k_{1}a)+
H_{0}^{(1)}(k_{0}a)J_{1}(k_{1}a)}~,
\end{equation}
\begin{multline}\label{8.5}
C_{\pm 1}=
\\
\gamma_{1}\frac{J_{1}(k_{0}a)\left [
J_{0}(k_{1}a)-(k_{1}a)^{-1}J_{1}(k_{1}a)\right ] -\beta
J_{1}(k_{1}a)\left [
J_{0}(k_{0}a)-(k_{0}a)^{-1}J_{1}(k_{0}a)\right ] } {\beta
J_{1}(k_{1}a)\left [
H^{(1)}_{0}(k_{0}a)-(k_{0}a)^{-1}H^{(1)}_{1}(k_{0}a)\right ] -
H^{(1)}_{1}(k_{0}a)\left [
J_{0}(k_{1}a)-(k_{1}a)^{-1}J_{1}(k_{0}a)\right ] }~,
\end{multline}
etc.

We employ the notation:
\begin{equation}\label{8.6}
\zeta:=k_{1}/k_{0}~,~~\delta:=k_{0}a~.
\end{equation}
Due to the hypothesis of low frequencies (and/or small cylinder
radius),
\begin{equation}\label{8.8}
\delta<<1~.
\end{equation}
We employ the small-argument asymptotic forms of the Bessel and
Neumann functions \cite{AbSt68}
\begin{equation}\label{8.10}
J_{0}(\delta)\sim 1-\delta^{2}/4 ~~,~~J_{1}(\delta)\sim \delta/2
~~,~~Y_{0}(\delta)\sim (2/\pi)\ln\delta~~,~~Y_{1}(\delta)\sim
-2/\pi\delta~~;~~\delta\rightarrow 0~,
\end{equation}
to find
\begin{equation}\label{8.17}
C_{0}\sim \tilde{C}_{0}:=
-\gamma_{0}\frac{i\pi\delta^{2}}{4\beta}(-\zeta+\beta)
~~;~~\delta\rightarrow 0~,
\end{equation}
\begin{equation}\label{8.24}
C_{\pm 1}\sim \tilde{C}_{\pm 1}:=
-\gamma_{1}\frac{i\pi\delta^{2}}{4}\frac{1-\zeta\beta}{1+\zeta\beta}
~~;~~\delta\rightarrow 0~.
\end{equation}
Thus, $C_{0}$ and $C_{\pm1}$ are $\mathcal{O}(\delta^{2})$ as
$\delta\rightarrow 0$. In the same way we show that $C_{|m|>1}$
vanishes faster than $\delta^{2}$, so that to second order in
$\delta$ we can write
\begin{equation}\label{9.1}
u_{0}^{d}(\mathbf{x},\omega)\sim\sum_{m=-1}^{1}
\tilde{C}_{m}H_{m}^{(1)}(k_{0}r)\exp\ ( im\theta\ )~~;~~\forall
\mathbf{x}\in \Omega_{0}~~;~~\delta\rightarrow 0~,
\end{equation}
or, in other terms,
\begin{equation}\label{9.3}
u_{0}^{d}(\mathbf{x},\omega)\sim
\frac{i\pi\delta^{2}}{4}\frac{\zeta-\beta}{\beta}H_{0}^{(1)}(k_{0}r)-
\frac{\pi\delta^{2}}{2}\frac{1-\zeta\beta}{1+\zeta\beta}H_{1}^{(1)}(k_{0}r)
\cos(\theta-\theta^{i}) ~~;~~\forall \mathbf{x}\in
\Omega_{0}~~;~~\delta\rightarrow 0~.
\end{equation}
It is customary, but not necessary, to measure the field in the
far-field zone. In this case, we employ the large-argument
asymptotic form of the Hankel functions \cite{AbSt68}:
\begin{equation}\label{9.4}
H_{n}^{(1)}(\xi)\sim \sqrt{\frac{2}{\pi\xi}}\exp\left [ i\left (
\xi-\frac{n\pi}{2}-\frac{\pi}{4}\right ) \right ]
~~;~~\xi\rightarrow \infty~,
\end{equation}
to find:
\begin{equation}\label{9.5}
u_{0}^{d}(\mathbf{x},\omega)\sim
\breve{u}_{0}^{d}(\theta,\theta^{i},\omega)\sqrt{\frac{2}{\pi
k_{0}r}}\exp\left [ i\left ( k_{0}r-\frac{\pi}{4}\right ) \right ]
~~;~~k_{0}r\rightarrow \infty~,
\end{equation}
wherein, the so-called {\it far-field scattering function} is
given (in the asymptotic low-frequency regime) by
\begin{equation}\label{9.6}
\breve{u}_{0}^{d}(\theta,\theta^{i},\omega)\sim\frac{i\pi\delta^{2}}{4}\left
[ \frac{\zeta-\beta}{\beta}-2\left (
\frac{\zeta\beta-1}{\zeta\beta+1}\right )
\cos(\theta-\theta^{i})\right ]~~;~~\delta\rightarrow 0~.
\end{equation}
Using (\ref{2.14}) we find that (\ref{9.3}) and (\ref{9.6})
become:
\begin{multline}\label{10.7}
u_{0}^{d}(\mathbf{x},\omega)\sim \frac{i\pi(k_{0}a)^{2}}{4}\left [
 \left ( \frac{\kappa_{1}}{\kappa_{0}}-1\right )
 H_{0}^{(1)}(k_{0}r)-
 2i\left ( \frac{\frac{\rho_{1}}{\rho_{0}}-1}
{\frac{\rho_{1}}{\rho_{0}}+1}\right )  H_{1}^{(1)}(k_{0}r)
\cos(\theta-\theta^{i})\right ]
\\
~~;~~\forall \mathbf{x}\in
\Omega_{0}~~;~~(k_{0}a)\rightarrow 0~,
\end{multline}
\begin{equation}\label{10.8}
\breve{u}_{0}^{d}(\theta,\theta^{i},\omega)\sim\frac{i\pi(k_{0}a)^{2}}{4}\left
[ \left ( \frac{\kappa_{1}}{\kappa_{0}}-1\right ) -2\left (
\frac{\frac{\rho_{1}}{\rho_{0}}-1}
{\frac{\rho_{1}}{\rho_{0}}+1}\right )
\cos(\theta-\theta^{i})\right ]~~;~~\delta\rightarrow 0~.
\end{equation}
The general problem in (\ref{10.7}) and (\ref{10.8}) is to express
$A$ and $B$ in terms of $C(\theta)~;~\forall \theta\in [0,2\pi[$
knowing that
\begin{equation}\label{10.15}
C(\theta)=A+B\cos(\theta-\theta^{i})~~;~~\forall \theta\in
[0,2\pi[~.
\end{equation}
Then
\begin{equation}\label{10.16}
\int_{0}^{2\pi}C(\theta)\cos n\theta d\theta =A\int_{0}^{2\pi}\cos
n\theta d\theta+B\int_{0}^{2\pi}\cos(\theta-\theta^{i})\cos
n\theta d\theta~~;~~\forall n\in\mathbb{Z}~.
\end{equation}
from which we find:
\begin{equation}\label{10.18}
A=\frac{1}{2\pi}\int_{0}^{2\pi}C(\theta) d\theta ~~,~~
B=\frac{1}{\pi\cos \theta^{i}}\int_{0}^{2\pi}C(\theta)\cos \theta
d\theta~.
\end{equation}
Applied to (\ref{10.8}), this gives:
\begin{equation}\label{12.2}
\kappa_{1}=\kappa_{0}\left [ 1
+\frac{4}{i\pi(k_{0}a)^{2}}\frac{1}{2\pi}\int_{0}^{2\pi}
\breve{u}_{0}^{d}(\theta,\theta^{i},\omega) d\theta \right ] ~,
\end{equation}
\begin{equation}\label{12.4}
\rho_{1}=\rho_{0}\left [ \frac{1-
\frac{2}{i\pi(k_{0}a)^{2}}\frac{1}{\pi\cos
\theta^{i}}\int_{0}^{2\pi}\breve{u}_{0}^{d}(\theta,\theta^{i},\omega)\cos
\theta d\theta}{1+ \frac{2}{i\pi(k_{0}a)^{2}}\frac{1}{\pi\cos
\theta^{i}}\int_{0}^{2\pi}\breve{u}_{0}^{d}(\theta,\theta^{i},\omega)\cos
\theta d\theta}\right ]~.
\end{equation}
This shows that: i) $\kappa_{1}$ can be retrieved independently of
$\rho_{1}$, ii) $\kappa_{1}$ is a linear function of the measured
scattered field, whereas iii) $\rho_{1}$ is a nonlinear function
of this field. More importantly: (\ref{12.2})-(\ref{12.4})
constitute a method for determining $\rho_{1}$ and the real and
imaginary parts of $\kappa_{1}$ from the (far-field) scattering
function, in an explicit, analytic manner.
\section{Numerical results}\label{s13}
We applied formulas (\ref{12.2}) and (\ref{12.4}) to retrieve the
density and complex compressibility from the far-field scattering
function. More precisely, we computed the relative errors (see
figures 2-4):
\begin{equation}\label{13.1}
  \delta_{\rho_{1}}:=\left | \frac{\tilde{\rho_{1}}-
  \rho_{1}}{\rho_{1}}\right | ,~~
  \delta_{\Re\kappa_{1}}:=\left | \frac{\Re\tilde{\kappa}_{1}-
  \Re\kappa_{1}}{\Re\kappa_{1}}\right | ,~~
  \delta_{\Im\kappa_{1}}:=\left | \frac{\Im\tilde{\kappa}_{1}-
  \Im\kappa_{1}}{\Im\kappa_{1}}\right | ,~
\end{equation}
(wherein $\tilde{\rho}_{1}$ is the value of density obtained from
(\ref{12.4}), and $\Re\tilde{\kappa}_{1}$, $\Im\tilde{\kappa}_{1}$
the values of the real and imaginary parts of the complex
compressibility obtained from (\ref{12.2})) over a range of
frequencies corresponding to $10^{-5}\leq k_{0}a\leq 1.5$. The
far-field data was simulated using (\ref{6.1}) in which the lower
and upper limits of the series were replaced by -5 and +5
respectively and use was also made of (\ref{9.5}). The other
parameters involved in the production of this data were:
$\rho_{0}=1000~kg/m^{3}$, $c_{0}=1500~m/s$, $\rho_{1}=1200~
kg/m^{3}$, $c_{1}=2500+i250~m/s$, $\theta^{i}=0$, $a=1.0~m$.
\begin{figure}
[ht]
\begin{center}
  \includegraphics[scale=0.6] {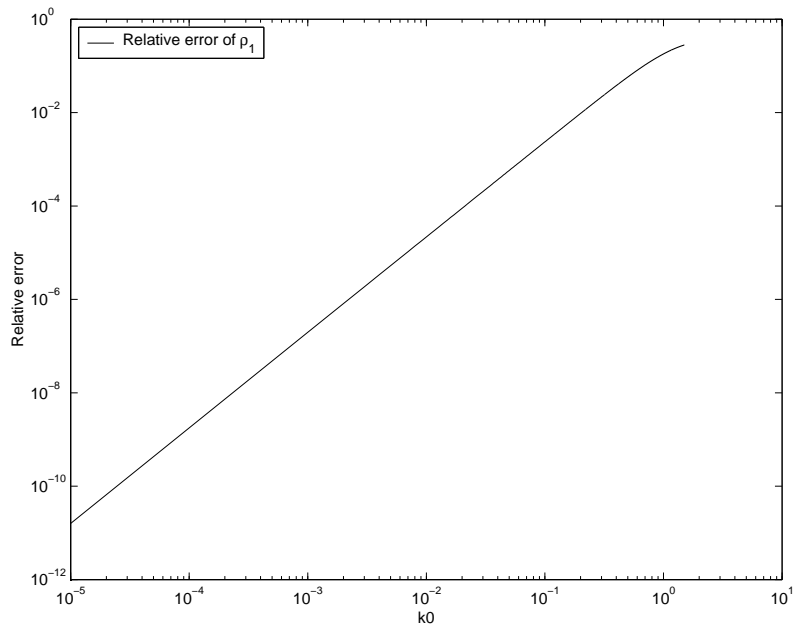}
  \caption{Relative error of  $\rho_{1}$}
  \label{rer}
  \end{center}
\end{figure}
\begin{figure}
[ht]
\begin{center}
  \includegraphics[scale=0.6] {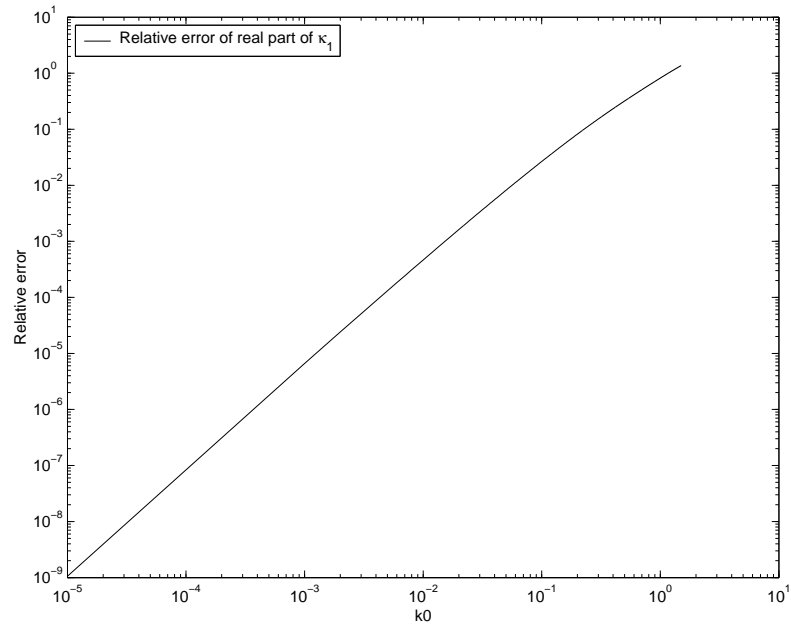}
  \caption{Relative error of $\Re\kappa_{1}$}
  \label{rerk}
  \end{center}
\end{figure}
\begin{figure}
[ht]
\begin{center}
  \includegraphics[scale=0.6] {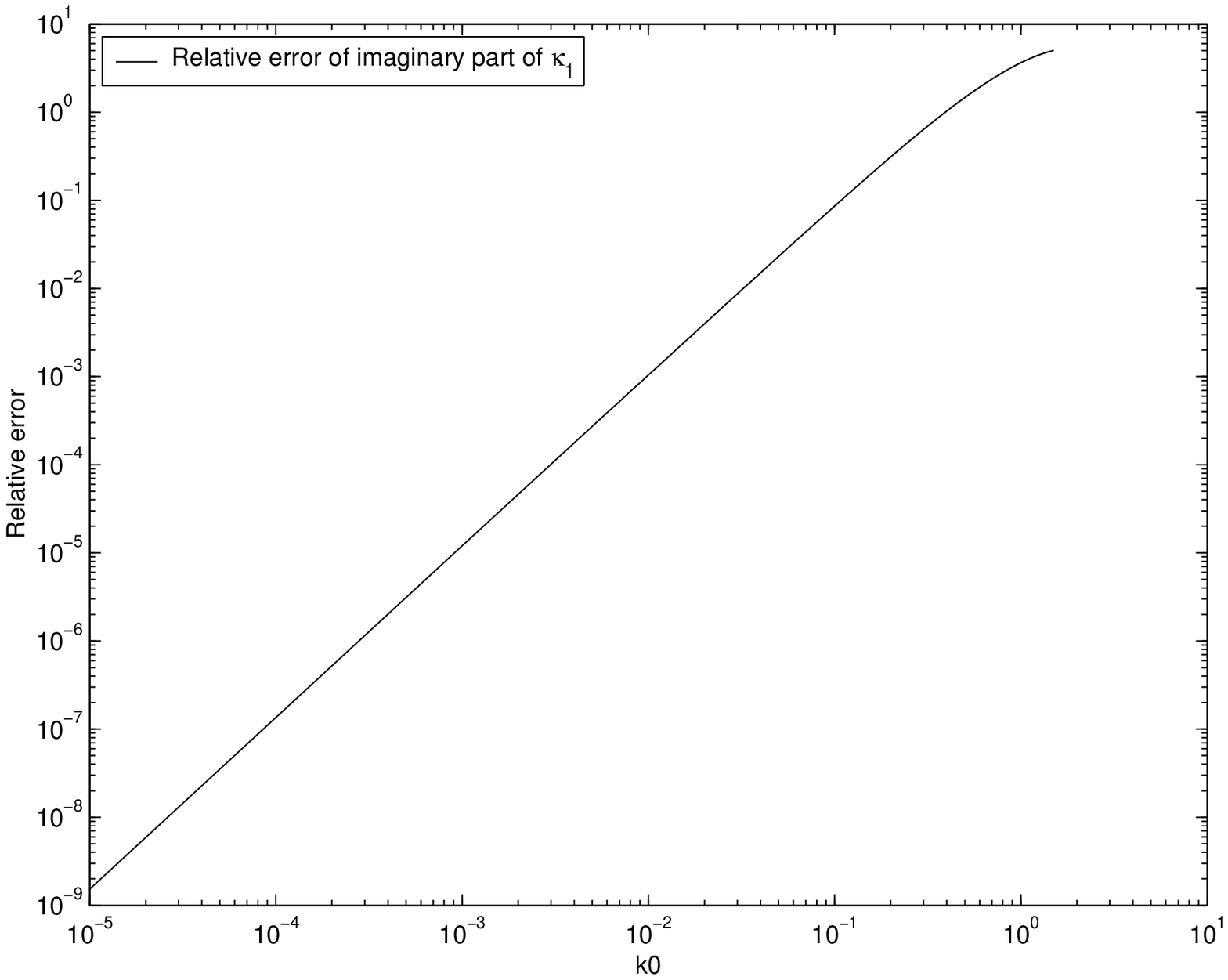}
  \caption{Relative error of  $\Im\kappa_{1}$}
  \label{reik}
  \end{center}
\end{figure}
One sees from figures 2-4 that in order to get relative errors of
the mechanical material parameters inferior or equal to 5\%, the
probe frequency should be such that $k_{0}a$ is not greater than
$\sim 0.1$.
\section{Conclusion}\label{s15}
Low frequency probe radiation is interesting in that it provides
solutions to the inverse medium problem which can be written in
{\it closed form} and which are {\it unique}. Moreover, these
solutions do not rely (as those appealing to the Born
approximation) on any assumptions concerning the compressibility
and density contrasts. Thus, the results presented herein are
valid for arbitrary values of these contrasts.

At the worst, the solutions of the inverse problem treated with
low-frequency probe radiation provide suitable starting solutions
for reconstructions carried out with higher-frequency probe
radiation as well as possible explanations of the difficulties
encountered in inverse medium problems such as the one considered
herein. They may also provide decent estimates of the material
parameters of  homogeneous (and even inhomogeneous) bodies of more
general shapes.

In case the characteristic dimension of the body (here the radius
$a$) is not known a priori, it can be determined from
high-frequency probe radiation using the asymptotic technique
described in \cite{Ho66}, \cite{LeBe96} \cite{MoFe53}.

The method outlined herein is obviously transposable to:
homogeneous fluid slabs and spheres, homogeneous elastic slabs,
circular cylinders and spheres, and to fluid-like or elastic
circular tubes and spherical shells.

\end{document}